\documentclass[preprint,onecolumn,nofootinbib]{revtex4}
\pdfoutput=1
\usepackage[colorlinks=true,linkcolor=blue,urlcolor=blue,filecolor=black,citecolor=red,pdfstartview=FitV,pdftitle={},pdfsubject={},pdfkeywords={},pdfpagemode=None,bookmarksopen=true]{hyperref}
\usepackage{graphicx}
\usepackage{amsmath}
\usepackage{amsfonts}
\usepackage{amssymb,ulem}
\usepackage{color}%
\usepackage{dcolumn}
\newcommand{\f}{\begin{equation}}
\newcommand{\ff}{\end{equation}}
\newcommand{\fa}{\begin{eqnarray}}
\newcommand{\ffa}{\end{eqnarray}}

\begin{document}
\title{Holographic superconductivity from higher derivative theory}
\author{Jian-Pin Wu $^{1}$}
\thanks{jianpinwu@mail.bnu.edu.cn}
\author{Peng Liu $^{2}$}
\thanks{phylp@jnu.edu.cn}
\affiliation{
$^1$ Institute of Gravitation and Cosmology, Department of
Physics, School of Mathematics and Physics, Bohai University, Jinzhou 121013, China}
\affiliation{$^{2}$ Department of Physics, Jinan University, Guangzhou 510632, China}
\begin{abstract}

We construct a $6$ derivative holographic superconductor model in the $4$-dimensional bulk spacetimes,
in which the normal state describes a quantum critical (QC) phase.
The phase diagram $(\gamma_1,\hat{T}_c)$ and the condensation as the function of temperature are worked out numerically.
We observe that with the decrease of the coupling parameter $\gamma_1$, the critical temperature $\hat{T}_c$ decreases and
the formation of charged scalar hair becomes harder.
We also calculate the optical conductivity.
An appealing characteristic is a wider extension of the superconducting energy gap,
comparing with that of $4$ derivative theory.
It is expected that this phenomena can be observed in the real materials of high temperature superconductor.
Also the Homes' law in our present models with $4$ and $6$ derivative corrections is explored.
We find that in certain range of parameters $\gamma$ and $\gamma_1$, the experimentally measured value of the universal constant $C$ in Homes' law can be obtained.

\end{abstract}
\maketitle
\tableofcontents
\section{Introduction}

The AdS/CFT correspondence \cite{Maldacena:1997re,Gubser:1998bc,Witten:1998qj,Aharony:1999ti}
provides a powerful tool to study the quantum critical (QC) dynamics,
which are described by CFT and are strongly coupled systems
without quasi-particles descriptions \cite{Sachdev:QPT}.
An interesting and important holographic QC dynamical system is that
a probe Maxwell field coupled to the Weyl tensor $C_{\mu\nu\rho\sigma}$
in the Schwarzschild-AdS (SS-AdS) black brane background,
which is a $4$ derivative theory and has been fully studied in
\cite{Myers:2010pk,Sachdev:2011wg,Hartnoll:2016apf,Ritz:2008kh,WitczakKrempa:2012gn,
WitczakKrempa:2013ht,Witczak-Krempa:2013nua,Witczak-Krempa:2013aea,Katz:2014rla,Bai:2013tfa}.
This system has zero charge density and can be understood as particle-hole symmetry.
Of particular interest is the non-trivial optical conductivity due to the introduction of Weyl tensor,
which is similar to the one in the superfluid-insulator quantum critical point (QCP) \cite{Myers:2010pk,Sachdev:2011wg,Hartnoll:2016apf}.
Further, higher derivative (HD) terms are introduced and the optical conductivity is studied in \cite{Witczak-Krempa:2013aea}.
They find that an arbitrarily sharp Drude-like peak can be observed at low frequency in the optical conductivity
and the bounds of conductivity found in \cite{Myers:2010pk,Ritz:2008kh} can be violated such that we have a zero DC conductivity at specific parameter.
Especially, its behavior resembles quite closely that of the $O(N)$ $NL\sigma M$ in the large-$N$ limit \cite{Damle:1997rxu}.
Therefore, the HD terms in SS-AdS black brane background provide possible route and alternative way to study the QC dynamics described by certain CFTs.

In this paper, we intend to study the holographic superconductor model in this holographic framework including HD terms.
The holographic superconductor model \cite{Hartnoll:2008vx,Hartnoll:2008kx,Horowitz:2010gk}
is an excellent example of the application of AdS/CFT in condensed matter theory (CMT),
which provides valuable lessons to access the high temperature superconductor in CMT.
In the original version of the holographic superconductor model \cite{Hartnoll:2008vx,Hartnoll:2008kx,Horowitz:2010gk},
the superconducting energy gap is $\omega_g/T_c\approx 8$.
This value is more than twice the one, which is $3.5$, in the weakly coupled BCS theory,
but roughly approximates that measured in high temperature superconductor materials \cite{Gomes:2007}.
Furthermore, by introducing $4$ derivative term based on Weyl tensor,
the holographic superconductor in the boundary theory dual to $5$ dimensional AdS black brane
is firstly constructed in \cite{Wu:2010vr}.
This model exhibits an appealing characteristic of
the extension of superconducting energy gap approximately varying from $6$ to $10$ \cite{Wu:2010vr}.
Next, lots of Weyl holographic superconductor models, including p wave and different backgrounds,
are constructed in \cite{Ma:2011zze,Momeni:2011ca,Momeni:2012ab,Momeni:2012uc,Roychowdhury:2012hp,Zhao:2012kp,Momeni:2013fma,Momeni:2014efa,Zhang:2015eea,Mansoori:2016zbp,Ling:2016lis}.
In particular, the extension of superconducting energy gap is also observed in \cite{Ma:2011zze,Mansoori:2016zbp}\footnote{In \cite{Momeni:2011ca,Momeni:2012ab,Momeni:2012uc,Roychowdhury:2012hp,Zhao:2012kp,Momeni:2013fma,Momeni:2014efa,Zhang:2015eea},
they study the s or p wave superconducting condensation from $4$ derivative term in $5$ dimensional AdS geometry.
But the computation of conductivity is absent.
In \cite{Mansoori:2016zbp}, the Weyl holographic superconductor in the $4$ dimensional Lifshitz black brane is explored
and the extension of the superconducting energy gap is also observed.
In addition, we would also like to point out that,
the running of superconducting energy gap is also observed in other holographic superconductor model from higher derivative gravity,
for example, the Gauss-Bonnet gravity \cite{Gregory:2009fj,Pan:2009xa} and the quasi-topological gravity \cite{Kuang:2010jc,Kuang:2011dy}.
But the value of $\omega_g/T_c$ is always greater than $8$.}.
Here, we extend the analysis to include the $6$ derivative term in the $4$ dimensional
bulk spacetime, in which the normal state describes a QC phase
and the electromagnetic (EM) self-duality loses.

We also study the Homes' law over our model.
Homes' law is an empirical law universally discovered in experiments of superconductors,
which states that the product of the DC conductivity $\sigma_{DC}(T_c)$ and the critical temperature $T_c$
has a linear relation to the superfluid density $\rho_S(T=0)$ at zero temperature.
Holographic investigation of Homes' law can be found in \cite{Erdmenger:2012ik,Erdmenger:2015qqa,Kim:2016jjk,Ling:2016lis}.
Our results show that the constant of the Homes' law can be observed in certain range of parameter $\gamma$ and $\gamma_1$ in our model,
which can be extended by adding additional structures to study the universal realization of the Homes' law in holography.

\section{Holographic framework}\label{sec-setup}

We shall construct a charged scalar hair black brane solution based SS-AdS black brane
\fa
\label{bl-br}
&&
ds^2=\frac{L^2}{u^2}\Big(-f(u)dt^2+dx^2+dy^2\Big)+\frac{L^2}{u^2f(u)}du^2\,,
\nonumber
\\
&&
f(u)=(1-u)p(u)\,,~~~~~~~
p(u)=u^2+u+1\,.
\ffa
$u=0$ is the asymptotically AdS boundary while the horizon locates at $u=1$.
The Hawking temperature of this system is
$
T=3/4\pi L^2
$.
And then, we introduce the actions for gauge field $A$ and charged complex scalar field $\Psi$
\begin{eqnarray}
&&
\label{action-SA}
S_A=\int d^4x\sqrt{-g}\Big(-\frac{L^2}{8g_F^2}F_{\mu\nu}X^{\mu\nu\rho\sigma}F_{\rho\sigma}\Big)\,,
\\
&&
\label{action-SPsi}
S_{\Psi}=\int d^4x\sqrt{-g}\Big(-|D_{\mu}\Psi|^2-m^2|\Psi|^2\Big)\,.
\end{eqnarray}
In the action $S_A$, $F=dA$ is the curvature of gauge field $A$ and the tensor $X$ is an infinite family of HD terms as \cite{Witczak-Krempa:2013aea}
\fa
X_{\mu\nu}^{\ \ \rho\sigma}&=&
I_{\mu\nu}^{\ \ \rho\sigma}-8\gamma_{1,1}L^2 C_{\mu\nu}^{\ \ \rho\sigma}
-4L^4\gamma_{2,1}C^2I_{\mu\nu}^{\ \ \rho\sigma}
-8L^4\gamma_{2,2}C_{\mu\nu}^{\ \ \alpha\beta}C_{\alpha\beta}^{\ \ \rho\sigma}
\nonumber
\\
&&
-4L^6\gamma_{3,1}C^3I_{\mu\nu}^{\ \ \rho\sigma}
-8L^6\gamma_{3,2}C^2C_{\mu\nu}^{\ \ \rho\sigma}
-8L^6\gamma_{3,3}C_{\mu\nu}^{\ \ \alpha_1\beta_1}C_{\alpha_1\beta_1}^{\ \ \ \alpha_2\beta_2}C_{\alpha_2\beta_2}^{\ \ \ \rho\sigma}
+\ldots
\,,
\label{X-tensor}
\ffa
where $I_{\mu\nu}^{\ \ \rho\sigma}=\delta_{\mu}^{\ \rho}\delta_{\nu}^{\ \sigma}-\delta_{\mu}^{\ \sigma}\delta_{\nu}^{\ \rho}$
is an identity matrix and $C^n=C_{\mu\nu}^{\ \ \alpha_1\beta_1}C_{\alpha_1\beta_1}^{\ \ \ \alpha_2\beta_2}\ldots C_{\alpha_{n-1}\beta_{n-1}}^{\ \ \ \mu\nu}$.
In the above equations (\ref{action-SA}) and (\ref{X-tensor}), we have introduced the factor of $L$ so that the coupling parameters $g_F$ and $\gamma_{i,j}$ are dimensionless.
But for later convenience, we shall work in units where $L=1$ in what follows.
Notice that we shall set $g_F=1$ in the following numerical calculation.
When $X_{\mu\nu}^{\ \ \rho\sigma}=I_{\mu\nu}^{\ \ \rho\sigma}$, the action $S_A$ reduces to the standard version of Maxwell theory.
For convenience, we denote $\gamma_{1,1}=\gamma$ and $\gamma_{2,i}=\gamma_i (i=1,2)$.
In this paper, we mainly focus on the $6$ derivative terms, i.e., $\gamma_1$ and $\gamma_2$ terms.
But since the effect of $\gamma_1$ and $\gamma_2$ terms is similar,
we only turn on $\gamma_1$ term through this paper.
Note that when other parameters are turned off, $\gamma_1$ is constrained in the region $\gamma_1\leq 1/48$ in SS-AdS black brane background \cite{Witczak-Krempa:2013aea}.
The upper bound of $\gamma_1$ is because of the requirement that the DC conductivity in the boundary theory
is positive \cite{Witczak-Krempa:2013aea}.

The action $S_{\Psi}$ supports a superconducting black brane \cite{Hartnoll:2008vx}.
$\Psi$ is the charged complex scalar field with mass $m$ and the charge $q$ of the Maxwell field $A$,
which can be written as $\Psi=\psi e^{i\theta}$ with $\psi$ being a real scalar field and $\theta$ a St\"uckelberg field.
$D_{\mu}=\partial_{\mu}-iqA_{\mu}$ is the covariant derivative.
For convenience, we choose the gauge $\theta=0$ and then, the EOMs of gauge field and scalar field can be derived as
\begin{eqnarray}
\label{eom-gauge-psi}
\nabla_{\nu}\big(X^{\mu\nu\rho\sigma}F_{\rho\sigma}\big)
-4q^2A^{\mu}\psi^2
=0\,,
\,\,\,\,\,\,\,\,\,\,\,\,\,
\big[\nabla^2-(m^2+q^2A^2)\big]\psi=0\,.
\end{eqnarray}

\section{Condensation}\label{sec-con-6}

In this section, we numerically construct a charged scalar hair black brane solution
with $6$ derivative term\footnote{As far as we know, the studies on $4$ derivative holographic superconductor
in the existing literatures except \cite{Mansoori:2016zbp} are focus on the
$5$ dimensional bulk spacetime. Though the qualitative properties
are expected to be similar between the $4$ and $5$ dimensional holographic superconductor
with $4$ derivative term, we also present the main properties in Appendix \ref{sec_HS_4} so that the paper is self-contained.}
and study its superconducting phase transition.
The ansatz for the scalar field and gauge field is taken as
\fa
\psi=\psi(u)\,,\,\,\,\,\,
A=\mu(1-u)a(u) dt\,,
\label{At}
\ffa
where $\mu$ is the chemical potential of the dual field theory.
The background EOMs can be explicitly written as
\fa
&&
\label{psi-eom-b}
\psi''+\Big(\frac{f'}{f}-\frac{2}{u}\Big)\psi'-\frac{m^2}{u^2f}\psi-q^2\frac{(u-1)^2\mu^2}{f^2}a^2\psi=0\,,
\
\\
&&
\label{a-eom-b}
a''+\Big(\frac{X'_3}{X_3}+\frac{2}{u-1}\Big)a'+\Big(\frac{X'_3}{(u-1)X_3}-\frac{2q^2\psi^2}{u^2fX_3}\Big)a=0\,,
\ffa
where the prime represents the derivative with respect to $u$ and for convenience,
the tensor $X_{\mu\nu}^{\ \ \rho\sigma}$ is denoted as
$
X_{A}^{\ B}=\{X_1(u),X_2(u),X_3(u),X_4(u),X_5(u),X_6(u)\}
$
with
$
A,B\in\{tx,ty,tu,xy,xu,yu\}
$.
This system is characterized by the dimensionless quantities $\hat{T}\equiv T/\mu$.
Without loss of generality, we set $m^2=-2$ in what follows.
It is easy to see that the asymptotical behavior of $\psi$ at the boundary $u=0$ is
\fa
\label{asy-psi}
\psi=u\psi_1+u^2\psi_2\,.
\ffa
Here, we treat $\psi_1$ as the source and $\psi_2$ as the expectation value.
And then, we set $\psi_1=0$ such that the condensate is not sourced.

\begin{figure}
\center{
\includegraphics[scale=0.5]{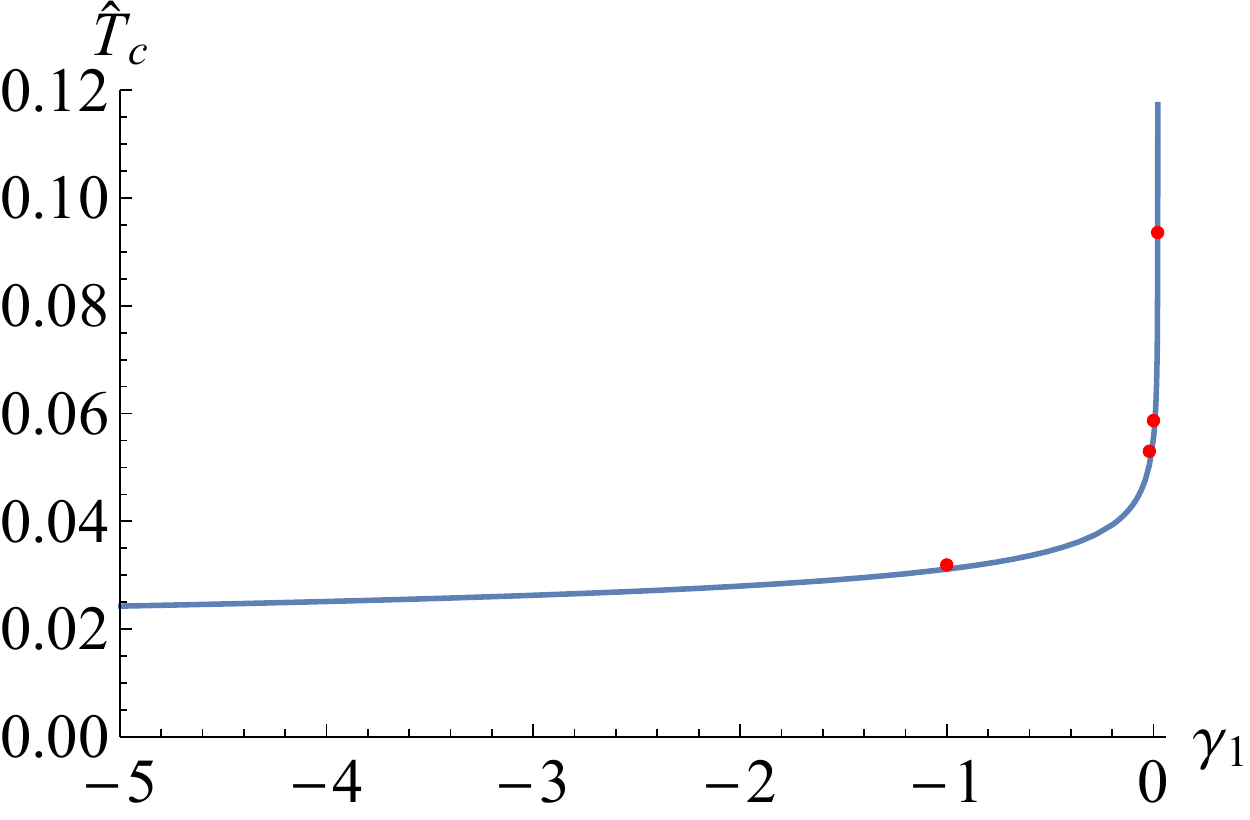}\ \hspace{0.5cm}
\includegraphics[scale=0.68]{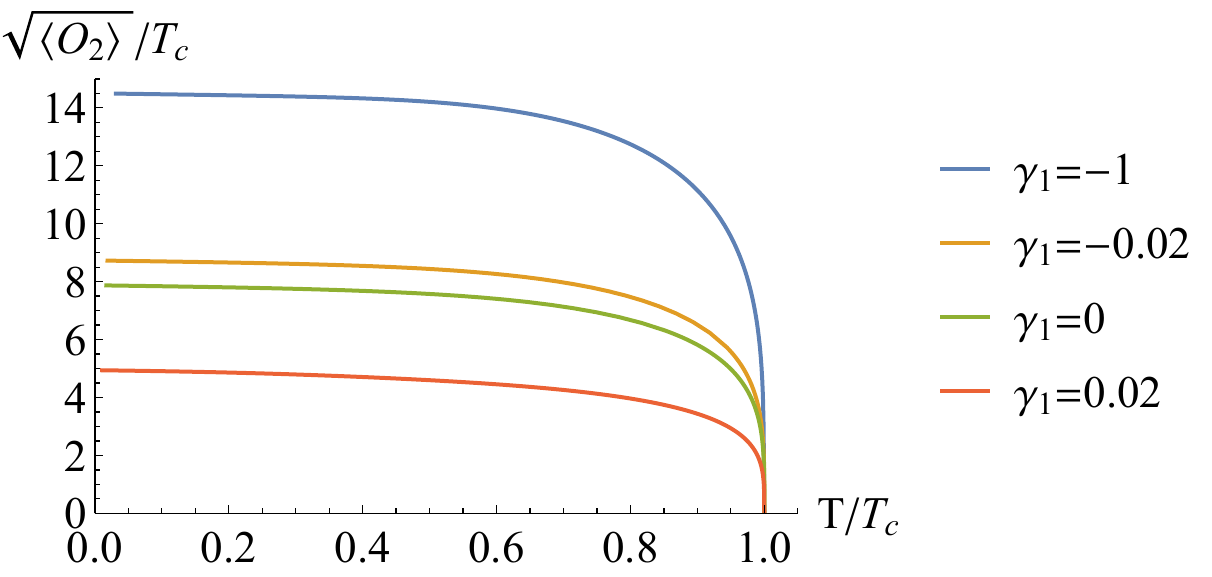}\ \\
\caption{\label{fig_con_6}Left plot: Phase diagram $(\gamma_1,\hat{T}_c)$ by finding static normalizable mode of charged scalar
field on the fixed background. The red points denote $\hat{T}_c$ for the representative $\gamma_1$,
which is obtained by fully solving the coupled EOMs (\ref{psi-eom-b}) and (\ref{a-eom-b}) numerically and listed in TABLE \ref{Tc_6}.
Right plot: The condensation as a function of temperature for representative $\gamma_1$.}}
\end{figure}

We firstly work out the phase diagram of the critical temperature $\hat{T}_c$ for the formation of the superconducting phase
as the function of the coupling parameter $\gamma_1$. It is convenient to estimate $\hat{T}_c$
by finding static normalizable mode of charged scalar field on the fixed background.
This method has been described detailedly in \cite{Horowitz:2013jaa,Ling:2014laa}.
Our result for the phase diagram $(\gamma_1,\hat{T}_c)$ is showed in the left plot in FIG.\ref{fig_con_6}.
The red points in this figure denote $\hat{T}_c$ for the representative $\gamma_1$,
which is obtained by fully solving the coupled EOMs (\ref{psi-eom-b}) as well as (\ref{a-eom-b})
and listed in TABLE \ref{Tc_6}.
We see that the $\hat{T}_c$ by finding static normalizable mode is in agreement with that shown in TABLE \ref{Tc_6}.
From the phase diagram $(\gamma_1,\hat{T}_c)$, we find that with the decrease of $\gamma_1$,
$\hat{T}_c$ decreases.
In particular, for the small $\gamma_1$, $\hat{T}_c$ seems to approach a fixed value.
Therefore, it seems reasonable to infer that the $6$ derivative term doesn't spoil the formation of charged scalar hair
even if the absolute value of $\gamma_1$ ($\gamma_1<0$) is large enough.
But note that the HD terms shall be introduced as a perturbative effect,
so we shall restrict $\gamma_1$ in a small region, $|\gamma_1|\ll 1$.
However, in order to see a more obvious effect from $6$ derivative term, we also relax the restriction of $\gamma_1$
and study the effect of $\gamma_1=-1$, in which the normal state has a sharp Drude peak and has been studied in \cite{Witczak-Krempa:2013aea}.

Now, we solve the coupling EOMs (\ref{psi-eom-b}) and (\ref{a-eom-b}) numerically and study the condensation behavior.
The result of the condensation $\sqrt{\langle O_2\rangle}/T_c$ as a function of the temperature $T/T_c$ is shown in the right plot in FIG.\ref{fig_con_6}.
We observe that with the decrease of $\gamma_1$, the condensation becomes much larger.
It implies a larger superconducting energy gap $\omega_g/T_c$, which is also observed in the following study of optical conductivity.
We also present the critical temperature $\hat{T}_c$ of the condensation phase transition with different $\gamma_1$
in TABLE \ref{Tc_6}. We see that when the $\gamma_1$ decreases, $\hat{T}_c$ goes down.
The tendency is consistent with that shown in the left plot in FIG.\ref{fig_con_6}.

\begin{widetext}
\begin{table}[ht]
\begin{center}
\begin{tabular}{|c|c|c|c|c|c|c|}
         \hline
~$\gamma_1$~ &~$-1$~&~$-0.02$~&~$0$~&~$0.02$~
          \\
        \hline
~$\hat{T}_c$~ & ~$0.0319$~&~$0.0530$~&~$0.0587$~ & ~$0.0936$~
          \\
        \hline
\end{tabular}
\caption{\label{Tc_6} The critical temperature $\hat{T}_c$ with different $\gamma_1$.}
\end{center}
\end{table}
\end{widetext}

\section{Superconductivity}\label{sec-supconductivity}

Given the charged scalar hair black brane solution from $6$ derivative term,
we can study the optical conductivity.
To this end, we turn on the perturbations of the gauge field along $x$ direction as
$\delta a_x(t,u)\sim\delta a_x(u)e^{-i\omega t}$.
And then the perturbative equation can be derived as
\fa
\label{eomp}
\delta a_x''
+\Big(\frac{X_5'}{X_5}+\frac{f'}{f}\Big)\delta a_x'
+\Big(\frac{\omega^2}{f^2}\frac{X_1}{X_5}-\frac{2q^2\psi^2}{u^2fX_5}\Big)\delta a_x=0\,.
\ffa
With the ingoing boundary condition at horizon,
we numerically solve the above perturbative equation
and read off the conductivity in terms of
\fa
\sigma(\omega)=\frac{\partial_u\delta a_x}{i\omega\delta a_x}\Big|_{u=0}\,.
\ffa

The normal state from $6$ derivative term has been studied in \cite{Witczak-Krempa:2013aea}.
For $\gamma_1=-1$, a sharp Drude peak emerges at small frequency in the conductivity.
At mediate frequency, the optical conductivity exhibits a gap and then at large frequency, it saturates to the value $\sigma=1$.
While for $\gamma_1=1/48$, the DC conductivity is zero and the low frequency conductivity displays a hard-gap-like behavior.
In addition, a pronounced peak emerges at mediate frequency.
As pointed out in \cite{Witczak-Krempa:2013aea}, these results are similar with that
of the large-$N$ $O(N)$ $NL\sigma M$ model and deserve further exploration \cite{Damle:1997rxu}.

\begin{figure}
\center{
\includegraphics[scale=0.55]{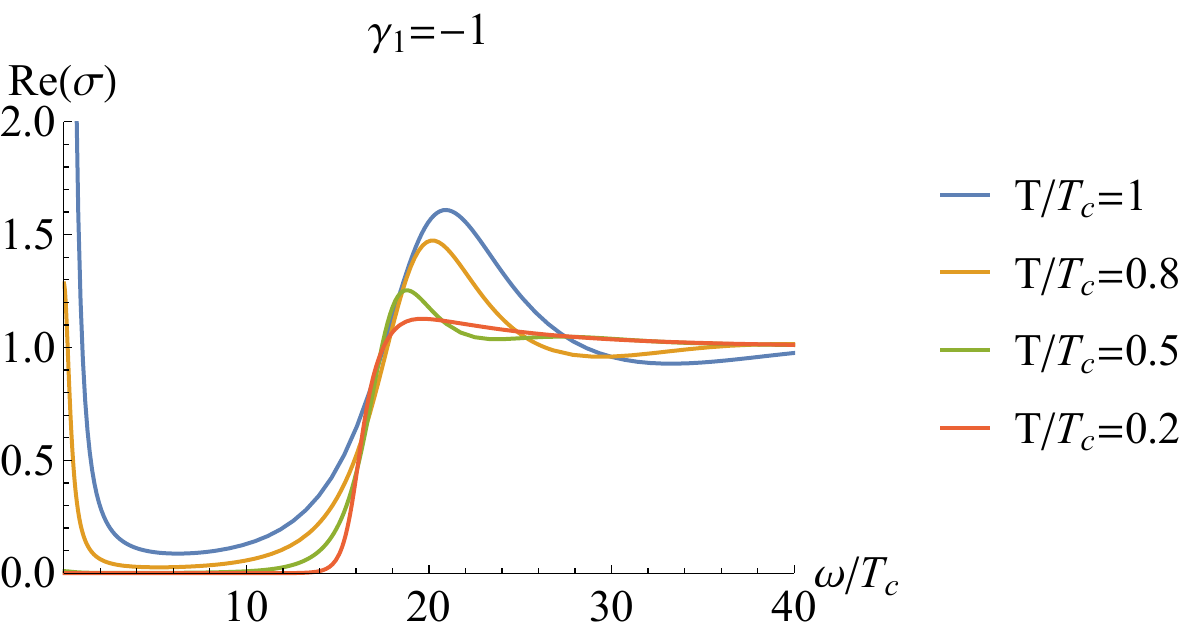}\ \hspace{0.5cm}
\includegraphics[scale=0.55]{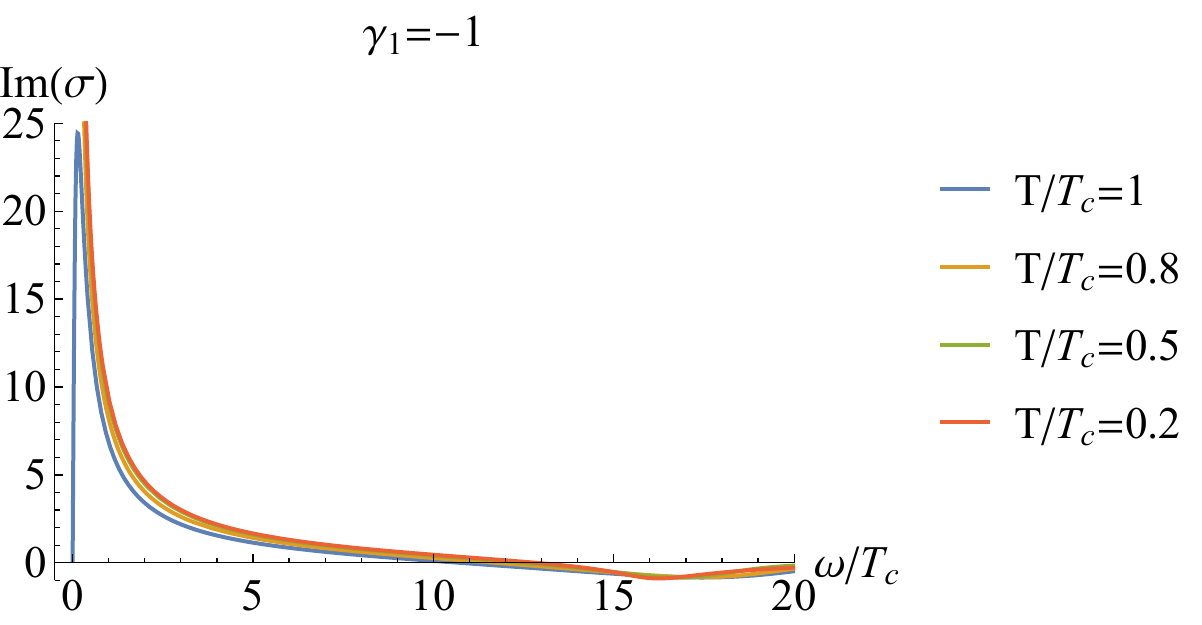}\ \\
\includegraphics[scale=0.55]{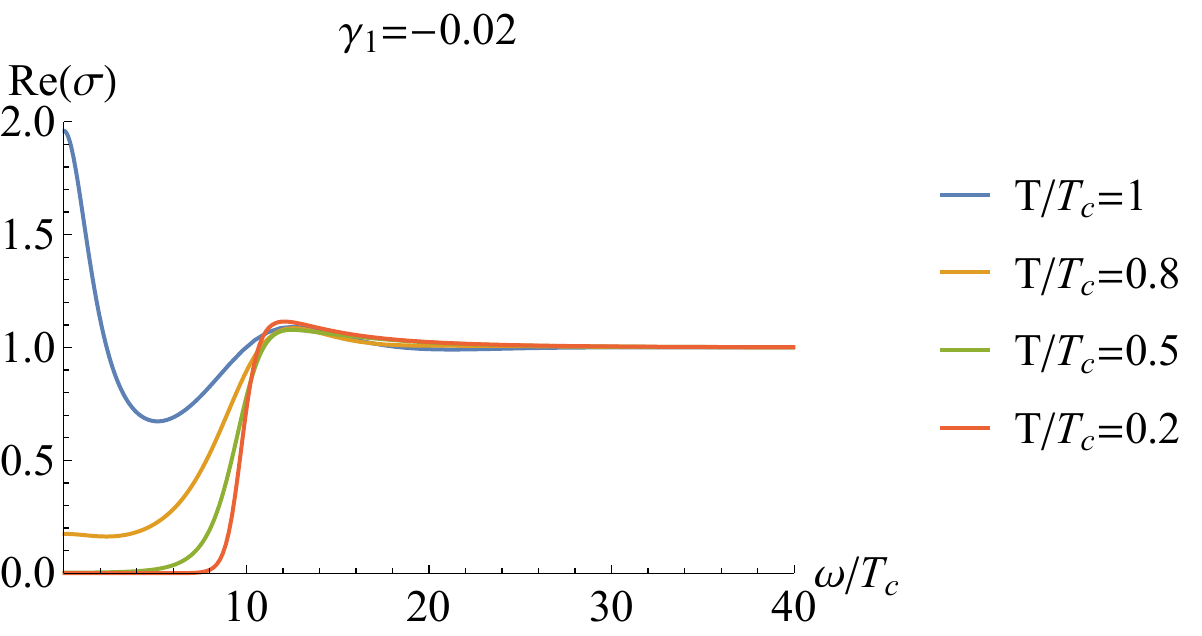}\ \hspace{0.5cm}
\includegraphics[scale=0.55]{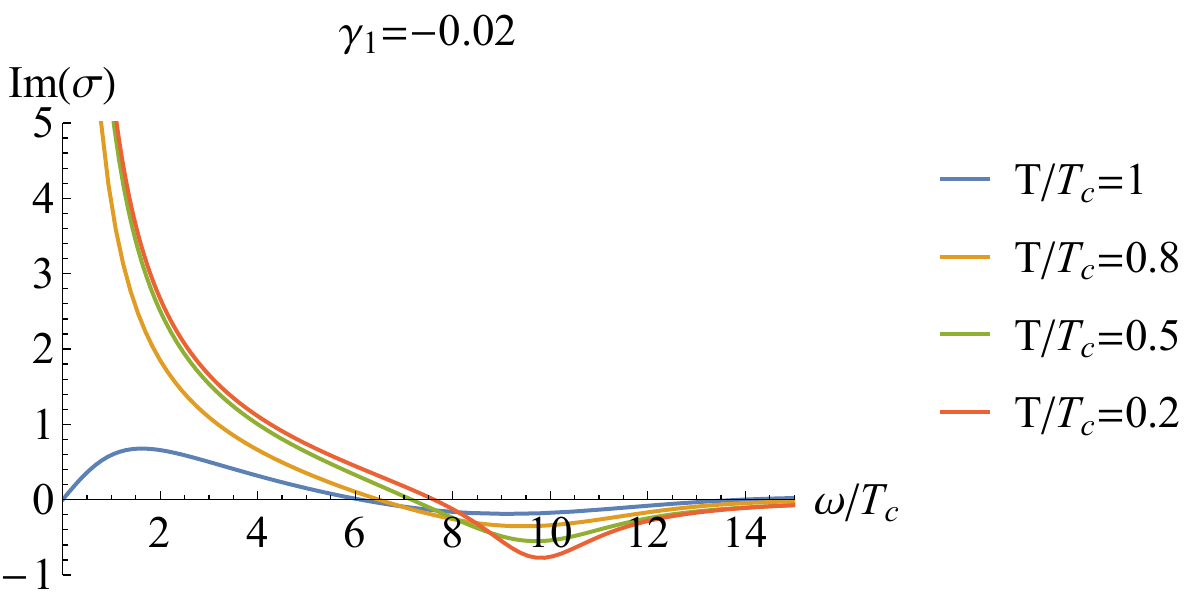}\ \\
\includegraphics[scale=0.55]{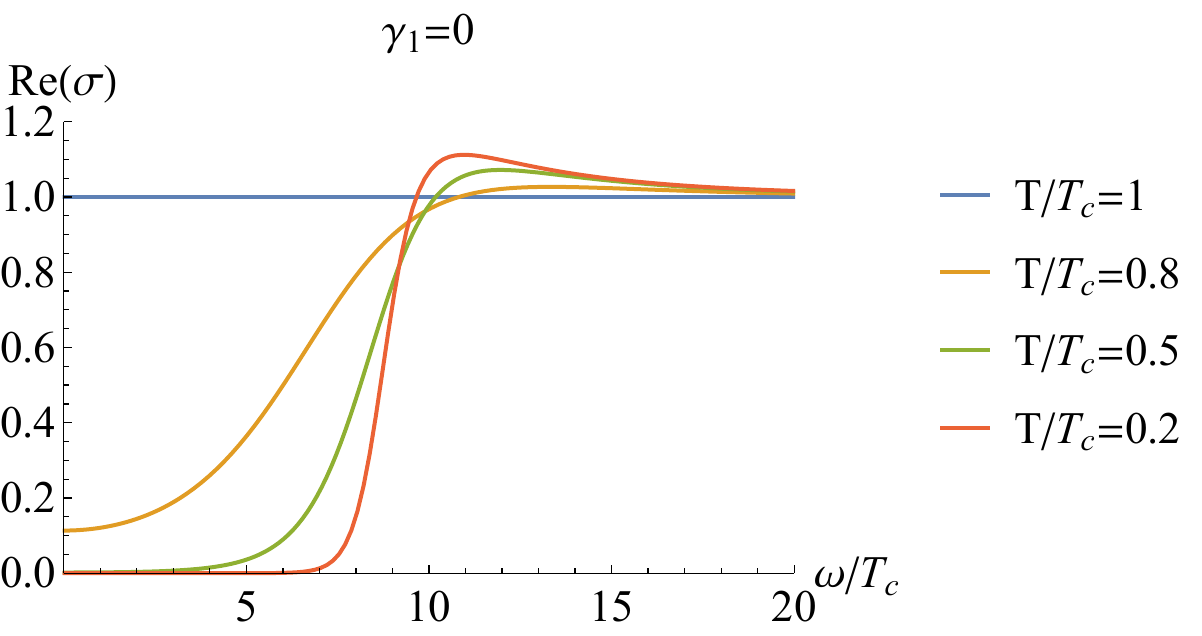}\ \hspace{0.5cm}
\includegraphics[scale=0.55]{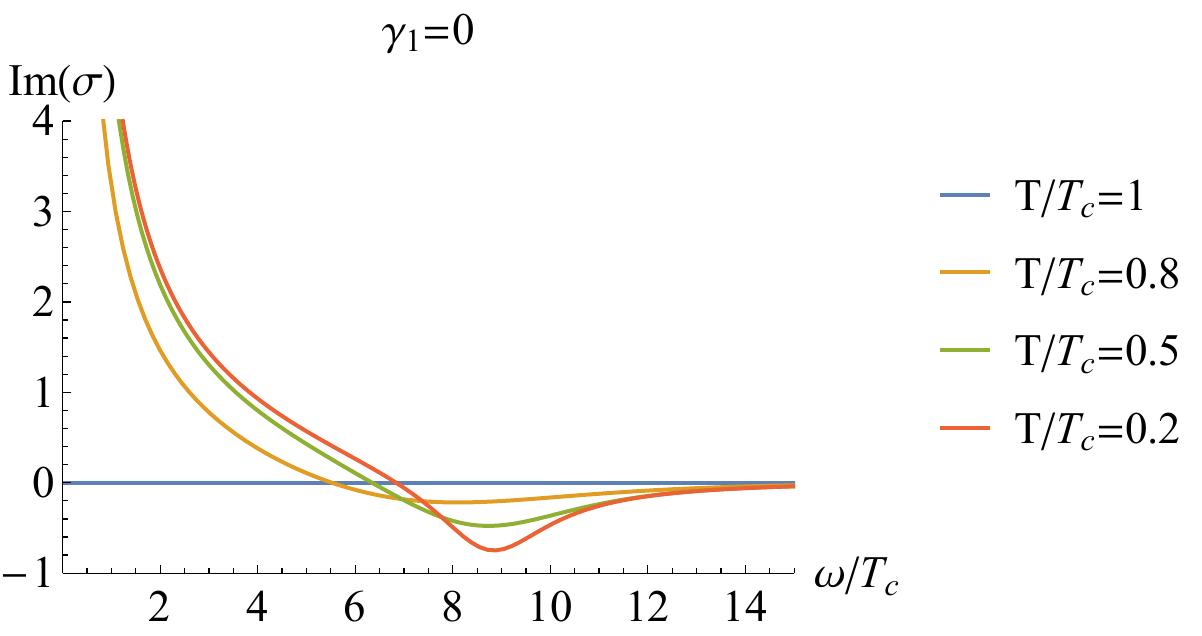}\ \\
\includegraphics[scale=0.55]{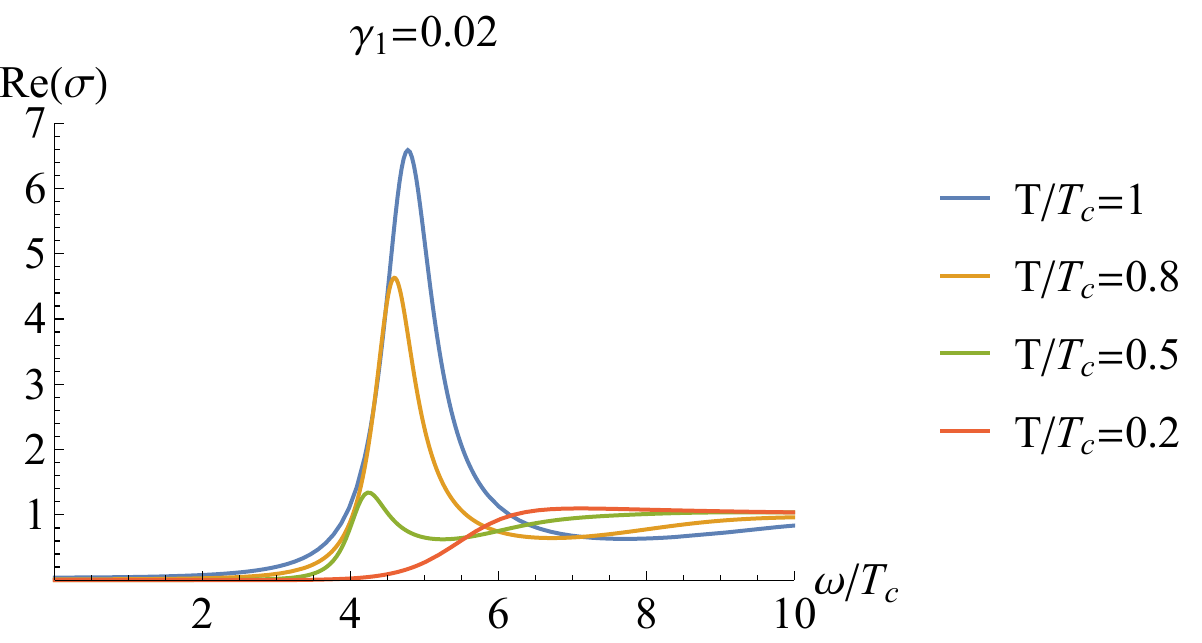}\ \hspace{0.5cm}
\includegraphics[scale=0.55]{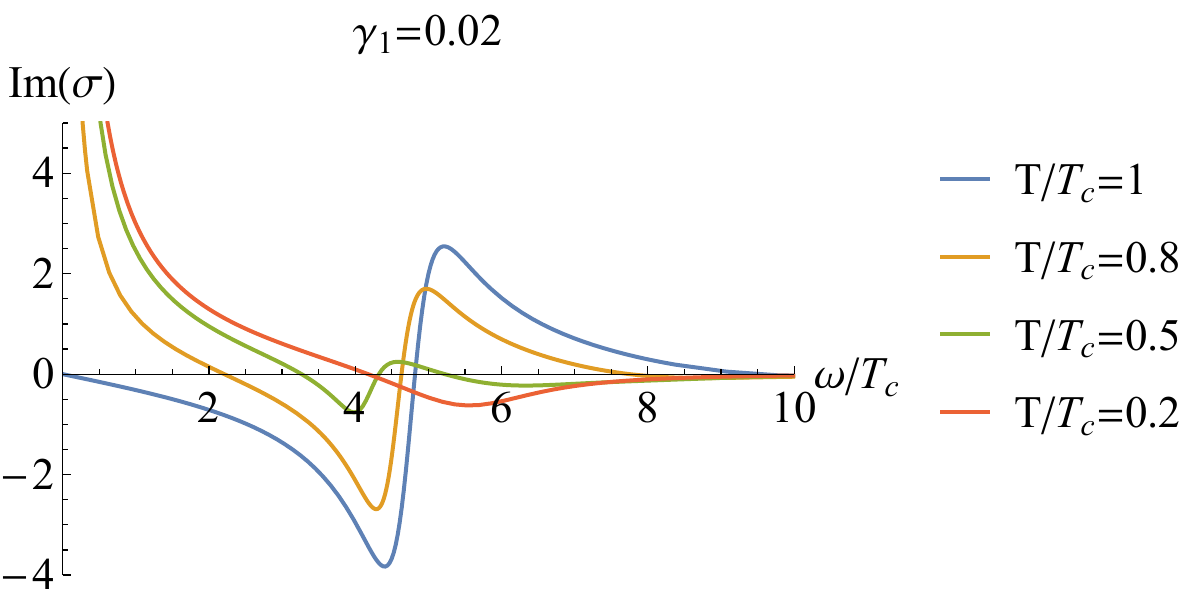}\ \\
\caption{\label{conductivity_6} Real and imaginary parts of conductivity as a function of the frequency with different $\gamma_1$.}}
\end{figure}

Now, we study the properties of the conductivity when the charged scalar hair is formed.
First, as the standard version of holographic superconductor model \cite{Hartnoll:2008vx},
a pole is observed at $\omega=0$ in the imaginary part of the conductivity (right plot in FIG.\ref{conductivity_6}),
which corresponds to a delta function at $\omega=0$ according to the Kramers-Kronig (KK) relation.
It indicates that the superconductivity emerges.
Subsequently, we mainly focus on two cases: one is for $\gamma_1=-1$ and another for $\gamma_1=0.02$.
For $\gamma_1=-1$, we notice that after the superconducting phase has formed,
the DC conductivity is still finite near the critical temperature (FIG.\ref{conductivity_6}).
It contributes to the normal component to the conductivity and means that our holographic model with $\gamma_1=-1$ is a two-fluid model.
With the decrease of the temperature, the DC conductivity goes down and finally vanishes at low temperature,
which means that normal component of the electron fluid is decreasing to form the superfluid component.
In addition, the soft gap at low frequency in normal state gradually becomes a superconducting gap with the decrease of the temperature.
While for $\gamma_1=0.02$, we find that with the decrease of the temperature,
the pronounced peak at the mediate frequency gradually decreases
and a supercondcting energy gap comes into being.

One important property of our present model is the extension of the superconducting energy gap,
which can be obtained by locating the minimal value of the imaginary part of the conductivity (see FIG.\ref{conductivity_6}).
The quantitative results are shown in TABLE \ref{Egap_6}\footnote{We adopted the convention that the unit of the system is the chemical potential here
as \cite{Ling:2014laa} but not the charge density as most of the present literatures. So the superconducting energy gap is
$\omega_g/T_c\simeq 9$ but not $\omega_g/\tilde{T}_c\simeq 8$.}.
We notice that comparing with the $4$ derivative model (see \cite{Wu:2010vr} and also Appendix \ref{sec_HS_4})
there is a wider extension of the energy gap, which ranges from $5.5$ to $16.2$ when $\gamma_1$ changes from $0.02$ to $-1$.
We expect that the extension of the energy gap in our holographic model has a corresponding partner in high temperature superconductor
and we pursuit it in future.

\begin{widetext}
\begin{table}[ht]
\begin{center}
\begin{tabular}{|c|c|c|c|c|c|c|}
         \hline
~$\gamma_1$~ &~$-1$~&~$-0.02$~&~$0$~&~$0.02$~
          \\
        \hline
~$\omega_g/T_c$~ & ~$16.253$~&~$9.886$~&~$8.880$~ & ~$5.529$~
          \\
        \hline
\end{tabular}
\caption{\label{Egap_6} The superconducting energy gap $\omega_g/T_c$ with different $\gamma_1$.}
\end{center}
\end{table}
\end{widetext}

\section{Homes' law}\label{sec-Homes}

In previous sections and the appendix we investigate the superconductivity in presence of $4$ derivative and $6$ derivative terms.
Next we intend to study the Homes' law \cite{Homes2004,Homes2005}, a universal relation observed in laboratory, in our model.
For a large class of superconductivity materials there exists an empirical law, i.e., Homes' law, regardless of the details of the materials,
\begin{equation}\label{homeslaw}
  \rho_s(\hat{T}=0) = C \sigma_{DC}(\hat{T}_c) \hat{T}_c,
\end{equation}
where $\sigma_{DC}(\hat{T}_c)$ is the DC conductivity at a temperature slightly above the critical temperature $\hat{T}_c$,
$\rho_s (\hat{T}=0)$ the density of the condensation and $C$ a constant.

The universal constant $C \simeq 4.4$ for ab-plane high temperature superconductors and clean BCS superconductors,
while for $c$-axis high temperature materials and BCS superconductors in dirty limit $C \simeq 8.1$.
The most recent results show that for organic superconductors the $C=4\pm 2.1$ \cite{Homes2013}\footnote{Note that all the data of $C$ in here
has been converted from the original experimental data into our holographic framework.}.
Theoretically, Zaanen proposed an explanation for the Homes' law of high temperature superconductors in terms of the Planckian dissipation \cite{Zaanen2004}.
The superfluid density $\rho_S = 4\pi n_S e^2/m_e$, which has dimension $[t]^{-2}$.
Meanwhile, DC conductivity $\sigma_{DC} = 4\pi n_N e^2 \tau/m_e$,
which has dimension $[t]^{-1}$ and $\tau$ is a timescale related to the dissipation.
To balance the Homes relation in dimension, the critical temperature $T_c$ has to be converted to the inverse of a timescale.
This timescale is naturally deduced from the temperature through the energy-time uncertainty relation $\tau \sim \hbar/k_B T$.
Therefore the Homes law is equivalent to that the time scale of the dissipation is as short as permitted by quantum physics,
namely the Placnkian dissipation \cite{Zaanen2004}.

It is desirable to test if the universal Homes' law would also emerge in holographic method.
Earlier literatures concerning Homes' relation in holographic framework include \cite{Erdmenger:2012ik,Erdmenger:2015qqa,Kim:2016jjk,Ling:2016lis}.
In \cite{Erdmenger:2015qqa,Kim:2016jjk} the holographic models with helical structures and Q-lattices structures have been studied respectively,
and Homes' relation has been found solid in narrow range of parameters. In \cite{Ling:2016lis} the axion model with Weyl correction have been studied,
and it is found that a generalized Homes' relation holds in a relatively large range of parameters,
\begin{equation}\label{ghr}
  \rho_s (\hat{T}=0) = C \sigma_{DC}(\hat{T}_c) \hat{T}_c + D\,,
\end{equation}
where $D$ is a constant.
Next we investigate Homes' relation in $4$ derivative and $6$ derivative theories respectively.

\begin{figure}
\center{
\includegraphics[scale=0.5]{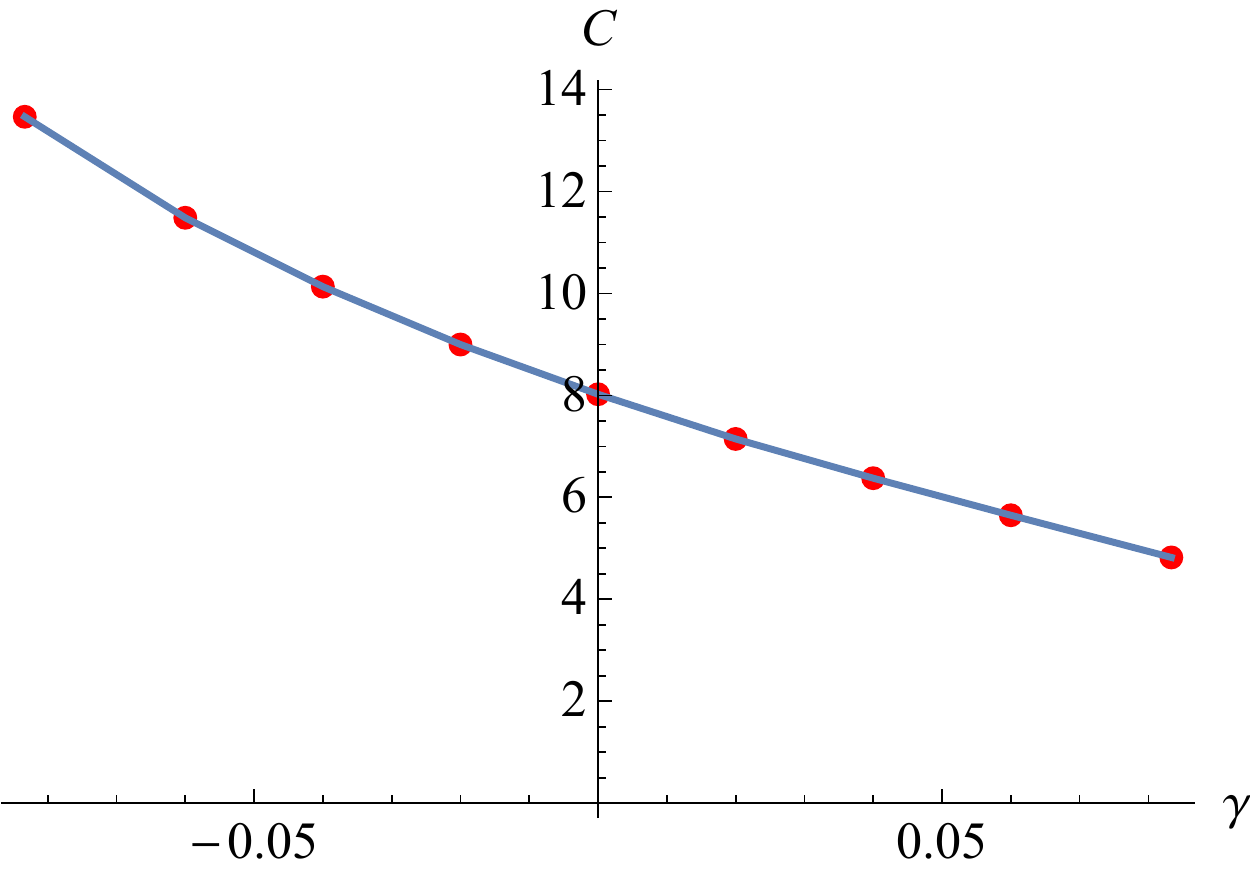}\ \hspace{0.5cm}
\includegraphics[scale=0.5]{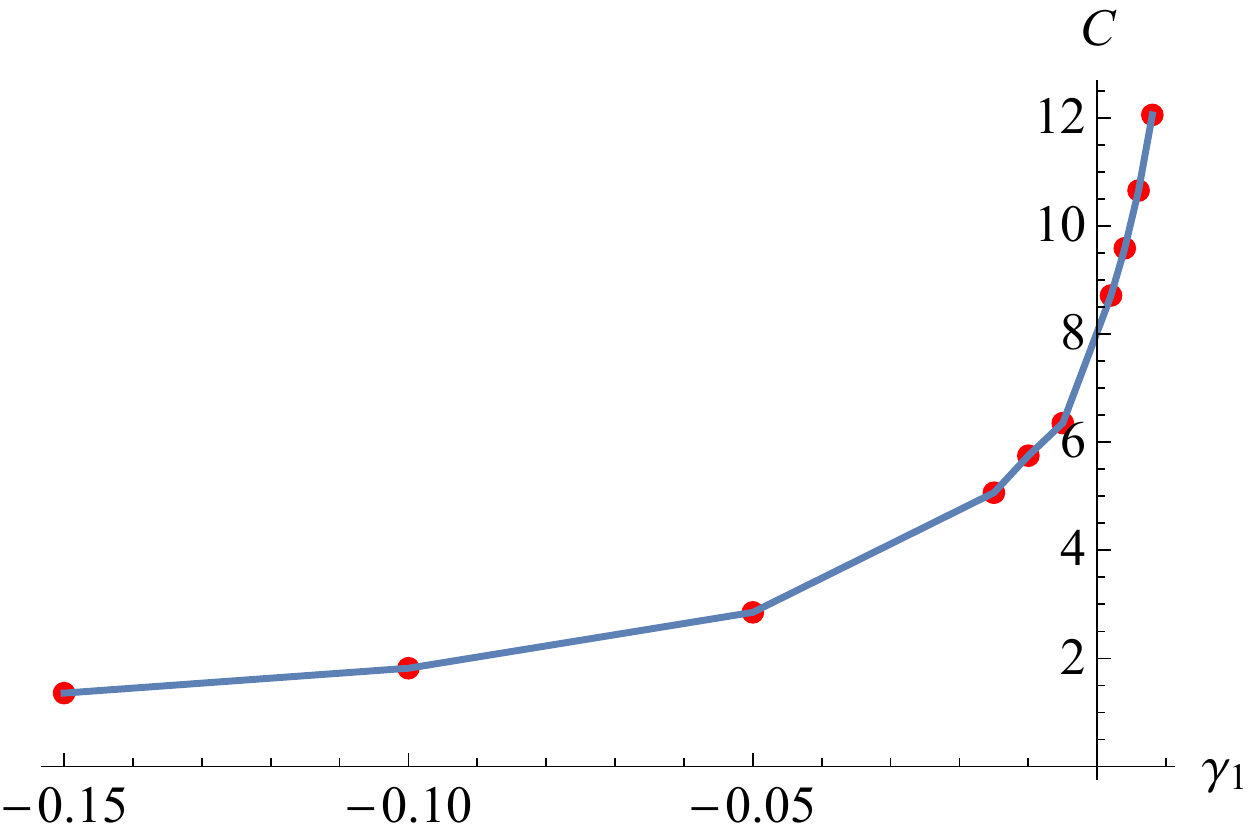}\ \\
\caption{\label{p4} Left: Homes' coefficient $C$ v.s. $\gamma$ in range $\gamma \in [-1/12,1/12]$ for $4$ derivative correction.
Right: Homes' coefficient $C$ v.s. $\gamma_1$ in range $\gamma_1 \in [-0.15,0.008]$ for $6$ derivative correction.}}
\end{figure}

The required ingredient to study Homes' law is $\rho_S(\hat{T}=0)$, $\hat{T}_c$ and $\sigma_{DC}(\hat{T}_c)$.
The computation of $T_c$ and $\sigma_{DC}(\hat{T}_c)$ is straightforward, while the superfluid density $\rho_S(\hat{T}=0)$
is identified as the coefficient of the pole in the imaginary part of the optical conductivity,
\begin{equation}\label{rhos_exp}
\text{Im}\, \sigma(\omega) = \frac{2\rho_S}{\pi \omega}.
\end{equation}
The constant $C$ v.s. $\gamma$ ($\gamma_1$) is shown in FIG. \ref{p4}.
It is interesting to notice that in both $4$ derivative and $6$ derivative theories, there exists a range of $\gamma$ ($\gamma_1$) where $C$ falls between the experimental results of Homes' relation. We expect the Homes' relation could be realized on these models when other structures, such as axions, lattices and dilatons are turned on.

\section{Conclusion and discussion}\label{sec-conclusion}

In this paper, we construct a $6$ derivative holographic superconductor model in the $4$-dimensional bulk spacetimes,
in which the normal state is a QC phase.
We find that with the decrease of the coupling parameter $\gamma_1$, the critical temperature $\hat{T}_c$ decreases and
the formation of charged scalar hair becomes harder.
But in terms of the present data we have obtained and the varying tendency of $\hat{T}_c$ with $\gamma_1$,
the $6$ derivative term doesn't seem to spoil the formation of the charged scalar hair and superconductivity.
And then, we study the optical conductivity.
One appealing property of our present model is the extension of the superconducting energy gap.
Comparing with that of $4$ derivative one, a wider extension of the energy gap, which ranges from $5.5$ to $16.2$ when $\gamma_1$ changes from $0.02$ to $-1$,
can be obtained. We expect that this phenomena can be observed in the real materials of high temperature superconductor
and we shall further explore it in future.
We also study the empirical and universal Homes' law in models with HD corrections.
The experimental results of Homes' law can be satisfied in certain range of parameters
both in $4$ derivative and $6$ derivative theories.

There are some interesting topics worthy of further investigation in future.
\begin{itemize}
  \item In our previous works \cite{Wu:2016jjd,Fu:2017oqa,Ling:2016dck}, the spatial linear dependent axions are incorporated into
  HD theory to study the optical conductivity \cite{Wu:2016jjd,Fu:2017oqa} and to implement metal-insulator transition \cite{Ling:2016dck}.
  We can incorporate more interesting structures, including axions \cite{Andrade:2013gsa}, Q-lattice \cite{Donos:2013eha},
  helical lattice \cite{Donos:2012js}, and the periodic lattice structure \cite{Horowitz:2012ky,Ling:2013nxa},
  into the HD theory to study the superconductivity, in particular the running of the energy gap, and the Homes' law in holography.
  \item We can construct the EM dual theory of the present model with HD terms and study its properties.
  Note that by introducing axion and dilaton fields, the EM duality for holographic p-wave superconductor has been explored in \cite{Gorsky:2017mkf}.
  \item It is valuable to study the transports and the quasi-normal modes in the superconducting state with HD terms at
  full momentum and energy spaces, which can provide far deeper insights into this holographic system than that at the zero momentum \cite{WitczakKrempa:2013ht,Amado:2009ts}.
\end{itemize}
We shall address these topics in future.

\begin{acknowledgments}
This work is supported by the Natural Science Foundation of China under Grant Nos. 11775036, 11305018,
and by the Natural Science Foundation of Liaoning Province under Grant No.201602013.

\end{acknowledgments}

\begin{appendix}

\section{Holographic superconductivity from $4$ derivatives}\label{sec_HS_4}

\begin{figure}
\center{
\includegraphics[scale=0.7]{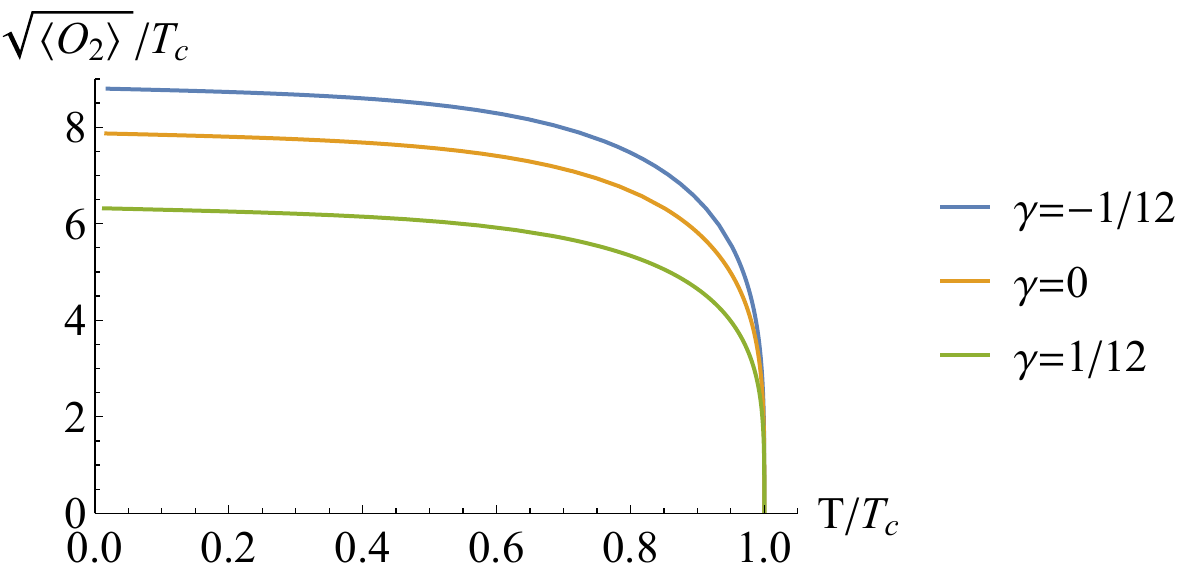}\ \\
\caption{\label{con_4} The condensation from $4$ derivative theory as a function of temperature for different $\gamma$.
}}
\end{figure}

In this Appendix, we present the main results of the holographic superconductivity from $4$ derivative term
in $4$ dimensional AdS geometry and give some comments.
FIG.\ref{con_4} exhibits the condensation $\sqrt{\langle O_2\rangle}/T_c$ as a function of temperature for various values of $\gamma$.
We see that with the decrease of $\gamma$, the condensation is formed harder,
which is consistent with the case of $5$ dimensional background \cite{Wu:2010vr}.
\begin{figure}
\center{
\includegraphics[scale=0.55]{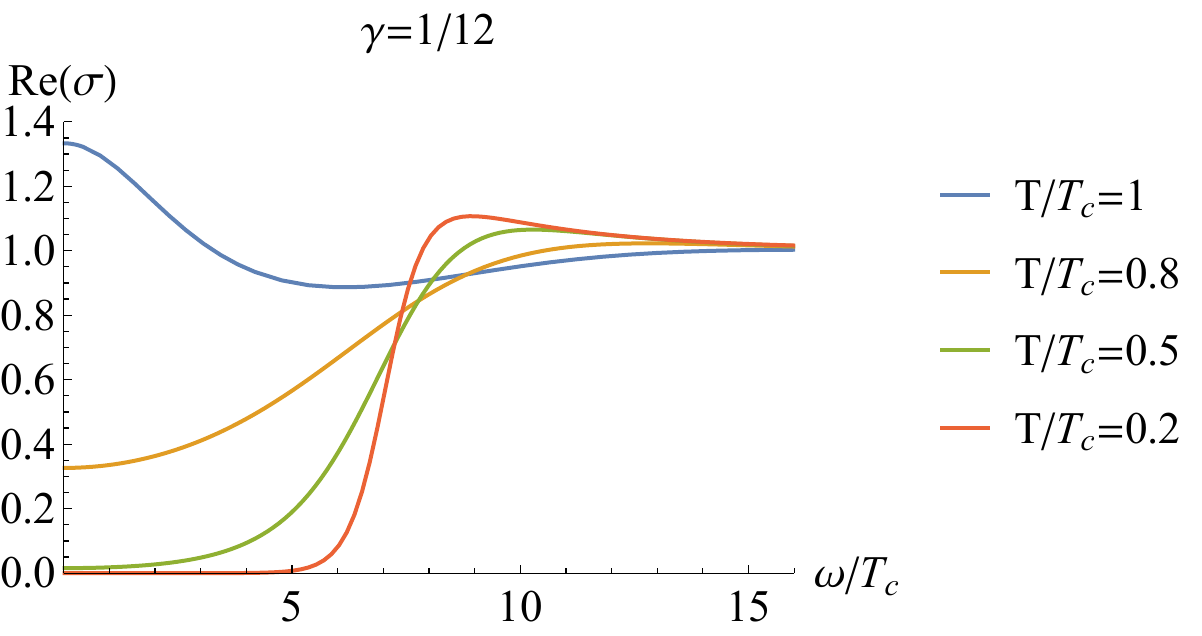}\ \hspace{0.5cm}
\includegraphics[scale=0.55]{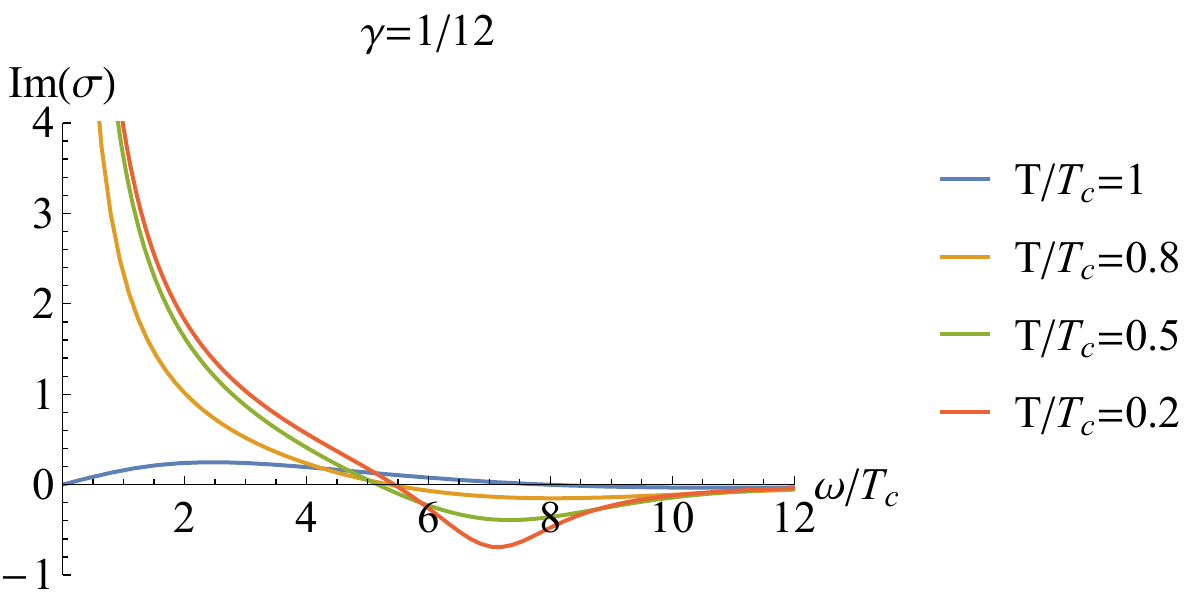}\ \\
\includegraphics[scale=0.55]{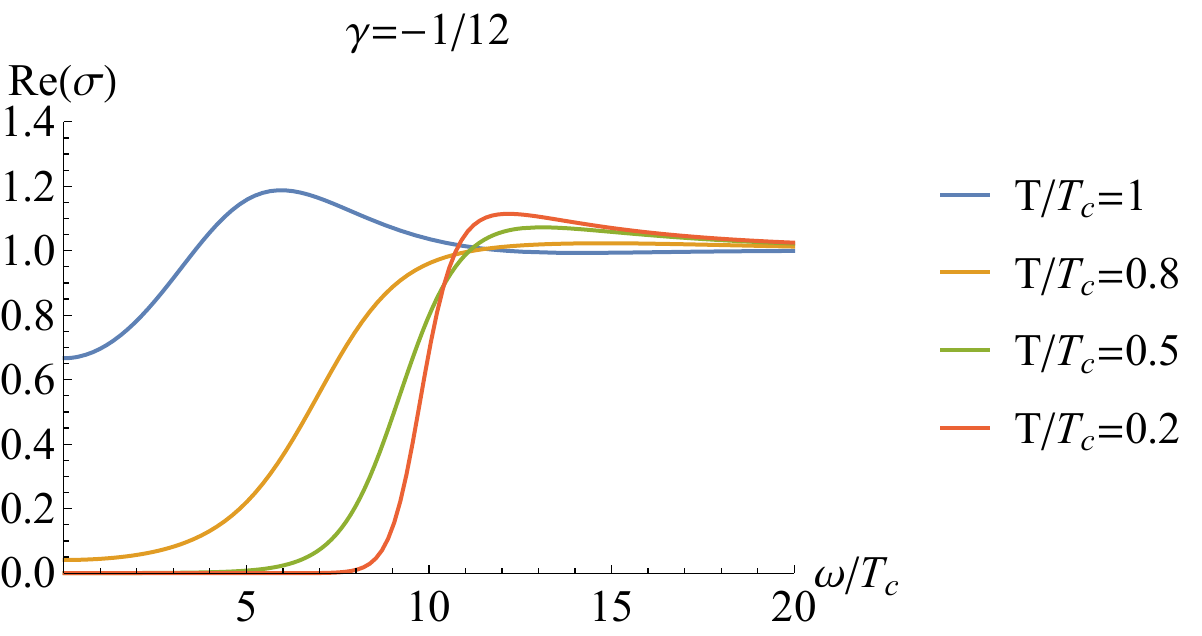}\ \hspace{0.5cm}
\includegraphics[scale=0.55]{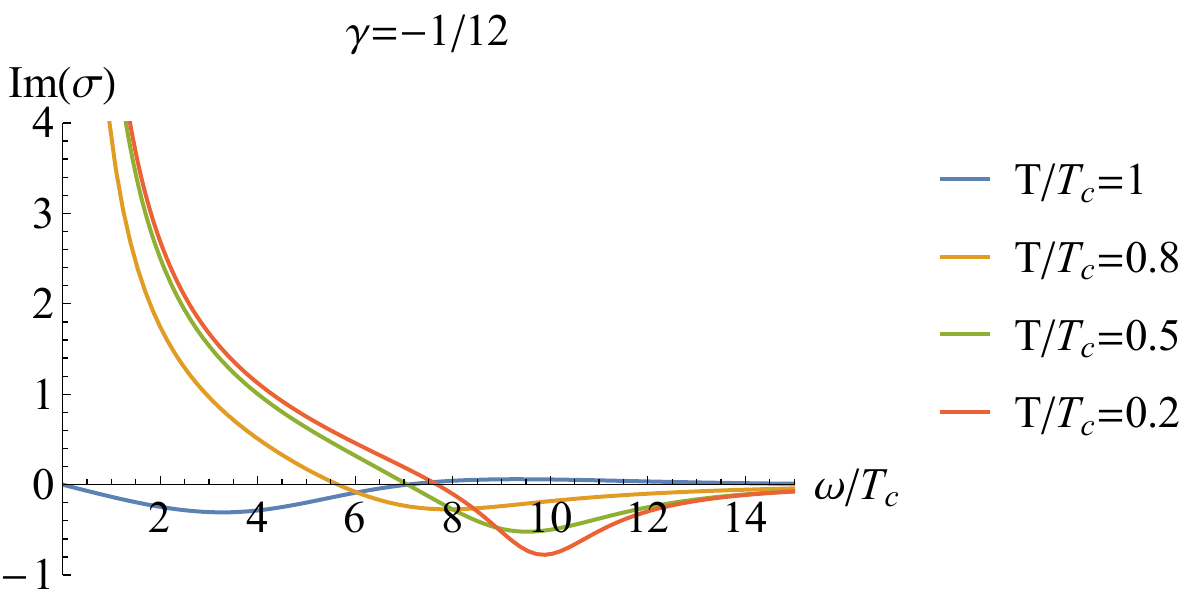}\ \\
\caption{\label{conductivity_4} Real and imaginary parts of conductivity as a function of the frequency for different $\gamma$.}}
\end{figure}

FIG.\ref{conductivity_4} shows the real and imaginary parts of the conductivity
as a function frequency for $\gamma=\pm1/12$, respectively.
The main properties are presented as follows.
\begin{itemize}
  \item \emph{Superconductivity.}
  The imaginary part of the conductivity exhibits a pole at $\omega=0$ (right plot in FIG.\ref{conductivity_4})
  when the temperature is below the critical temperature. It means that there
  is a delta function in the real part of the conductivity at $\omega=0$ and the superconductivity emerges.
  \item \emph{DC conductivity due to normal fluid.}
  The DC conductivity after the superconducting phase transition happens is still finite
  (left plot in FIG.\ref{conductivity_4}), which implies our model with $4$ derivative correction
  is a two-fluid model as the original version of holographic superconductor model \cite{Hartnoll:2008vx,Hartnoll:2008kx,Horowitz:2010gk}.
  But with the decrease of the temperature,
  the DC conductivity decreases and disappears at extremal low temperature, which implies that
  the normal component of the electron fluid is decreasing to form the superfluid component and disappears at extremal low temperature.
  \item \emph{Extension of energy gap.}
  From FIG.\ref{conductivity_4}, we see that there is a running of the energy gap when we tune the coupling parameter $\gamma$,
  which has been observed in the holographic superconductor model with $4$ derivative correction in $5$ dimensional AdS background \cite{Wu:2010vr}.
  The quantitative results are listed in TABLE\ref{EG_4}.
\begin{widetext}
\begin{table}[ht]
\begin{center}
\begin{tabular}{|c|c|c|c|c|c|c|}
         \hline
~$\gamma$~ &~$-1/12$~&~$0$~&~$1/12$~
          \\
        \hline
~$\omega_g/T_c$~ & ~$9.886$~&~$8.880$~&~$7.205$~
          \\
        \hline
\end{tabular}
\caption{\label{EG_4} The energy gap $\omega_g/T_c$ with different $\gamma$.}
\end{center}
\end{table}
\end{widetext}
\end{itemize}

\end{appendix}

\end{document}